\documentclass{article}
\usepackage{multirow}
\usepackage{graphicx}
     \PassOptionsToPackage{numbers, compress}{natbib}

\usepackage[preprint]{neurips_data_2024}

\usepackage[utf8]{inputenc} 
\usepackage[T1]{fontenc}    
\usepackage{hyperref}       
\usepackage{url}            
\usepackage{booktabs}       
\usepackage{amsfonts}       
\usepackage{nicefrac}       
\usepackage{microtype}      
\usepackage{xcolor}         
\usepackage{enumitem}
\usepackage{amsmath}
\usepackage{graphicx}
\usepackage{multirow}
\usepackage{adjustbox}
\usepackage{caption}

\title{MINT: a Multi-modal Image and Narrative Text Dubbing Dataset for Foley Audio Content Planning and Generation}

\author{%
Ruibo Fu$^{1\dagger}$ \quad Shuchen Shi$^{2\dagger}$ \quad Hongming Guo$^3$ \quad Tao Wang$^1$ \quad Chunyu Qiang$^{4,5}$\\ 
\textbf{Zhengqi Wen}$^1$ \quad \textbf{Jianhua Tao}$^{6,7}$ \quad \textbf{Xin Qi}$^{1,8}$ \quad \textbf{Yi Lu}$^{1,8}$ \quad \textbf{Xiaopeng Wang}$^{1,8}$\\
\textbf{Zhiyong Wang}$^{1,8}$ \quad \textbf{Yukun Liu}$^8$ \quad \textbf{Xuefei Liu}$^1$ \quad \textbf{Shuai Zhang}$^6$ \quad \textbf{Guanjun Li}$^1$\\
$^1$Institute of Automation, Chinese Academy of Sciences \quad $^2$Shanghai Polytechnic University \\ 
$^3$Beijing University of Posts and Telecommunications \quad $^4$Tianjin University \\
$^5$Kuaishou Technology Co., Ltd \quad $^6$Department of Automation, Tsinghua University \\
$^7$Beijing National Research Center for Information Science and Technology, Tsinghua University\\
$^8$University of Chinese Academy of Sciences \\
\texttt{ruibo.fu@nlpr.ia.ac.cn}
}

\begin{document}

\maketitle
\begin{abstract}
  Foley audio, critical for enhancing the immersive experience in multimedia content, faces significant challenges in the AI-generated content (AIGC) landscape. Despite advancements in AIGC technologies for text and image generation, the foley audio dubbing remains rudimentary due to difficulties in cross-modal scene matching and content correlation. Current text-to-audio technology, which relies on detailed and acoustically relevant textual descriptions, falls short in practical video dubbing applications. Existing datasets like AudioSet, AudioCaps, Clotho, Sound-of-Story, and WavCaps do not fully meet the requirements for real-world foley audio dubbing task. To address this, we introduce the Multi-modal Image and Narrative Text Dubbing Dataset (MINT), designed to enhance mainstream dubbing tasks such as literary story audiobooks dubbing, image/silent video dubbing. Besides, to address the limitations of existing TTA technology in understanding and planning complex prompts, a Foley Audio Content Planning, Generation, and Alignment (CPGA) framework is proposed, which includes a content planning module leveraging large language models for complex multi-modal prompts comprehension. Additionally, the training process is optimized using Proximal Policy Optimization based reinforcement learning, significantly improving the alignment and auditory realism of generated foley audio.  Experimental results demonstrate that our approach significantly advances the field of foley audio dubbing, providing robust solutions for the challenges of multi-modal dubbing. Even when utilizing the relatively lightweight GPT-2 model, our framework outperforms open-source multimodal large models such as LLaVA, DeepSeek-VL, and Moondream2. The dataset is available at \url{https://github.com/borisfrb/MINT}\footnote{The MINT dataset is licensed under CC BY-NC-SA-4.0 license.  $\dagger$ Equal contributions.}

\end{abstract}

\section{Introduction}  

Foley, or foley audio, refers to the generation of routine sound effects that are added to multimedia to enhance the immersion of experience \citep{forlysound1997}. In professional media production such as ﬁlms, animation, TV shows and short videos in social media, the foley audio is widely selected and fused into the corresponding video, which is called audio dubbing. Recently, with the rapid development and widespread application of AIGC technology in the entertainment sector, there is an increasing demand for the generation of highly immersive and realistic multimedia encompassing text, images, and audio, where foley audio dubbing is a very important but easily neglected part.

Currently, AIGC technology primarily focuses on the high-precision generation of information under the condition of precise content description, with most research centered on text and image modalities. Technologies like Sora excel in producing continuous visual content from minimal prompts, but generating foley audio remains rudimentary. Foley audio dubbing faces challenges in cross-modal scene matching and content correlation \citep{zhang2023complete}, requiring the alignment of audio descriptions with image and text modalities \citep{yang2020large}. In actual video dubbing scenarios, the most common form is to dub foley audio based on image and narrative text prompts, making it very challenging to generate matching and highly immersive audio. Hence, multi-modal dubbing is an urgently needed research direction.

The text-to-audio (TTA) technology \citep{xu2021text} is currently the mainstream solution for foley audio dubbing tasks. This technology necessitates that individuals with a certain level of relevant expertise provide a precise and fully acoustically relevant textual description, clear and suitable for the TTA model employed, to generate the desired scene's audio. Although multimodal promotes are utilized as input in tasks like image style transfer and text-based question-and-answer, textual promotes play an important role in AIGC tasks. When generating foley audio from textual promotes, the input must explicitly detail the sound events.

At present, research on TTA mainly focuses on two aspects: datasets and model architecture. In terms of datasets, the most classic dataset is AudioSet \citep{gemmeke2017audio}, which covers a wide range of sound events. To facilitate the modeling from text description to audio, the text descriptions have been carefully processed and annotated through various means. These text descriptions are very concise, containing only acoustic elements. However, some related datasets still differ from the requirements of actual video dubbing application scenarios in terms of modal type, description type, and audio quality. Examples of such datasets include AudioCaps \citep{kim2019audiocaps}, Clotho \citep{drossos2020clotho}, Sound-of-Story \citep{choi2021sound}, and WavCaps \citep{mei2023wavcaps}. These existing datasets are unable to fully meet the needs of foley audio dubbing applications in real-world scenarios. In terms of model architecture, AudioGen \citep{kreuk2022audiogen} uses a self-regressive TTA model. DiffSound \citep{yang2023diffsound} and Make-An-Audio \citep{huang2023make} employ discrete diffusion models. AudioLDM2 \citep{liu2023audioldm} and TANGO \citep{ghosal2023text} combine large language models (LLMs) and audio codecs. These improvements in model architecture mainly focus on accelerating the regeneration process and enhancing the quality of the generated foley audio. The generalization and robustness of models based on text inputs largely depend on whether the training data can cover the data distribution space of actual applications. Reinforcement learning has also been explored for TTA, offering potential advantages in optimizing models for perceptual audio quality through interaction-based learning \citep{majumder2024tango}. Unlike visual or textual data, audio quality is highly subjective and context-dependent, making it hard to design objective metrics.

However, in practical dubbing tasks, the prompts applicable for generating foley audio are typically multi-modal, and the text descriptions are often lengthy narratives rich in detail. While some research efforts aim to enhance the robustness of TTA models by using preprocessed text descriptions \citep{vyas2023audiobox}, most research primarily addresses issues of nonstandard and ambiguous descriptions. The modeling primarily focuses on descriptions highly related to acoustics, resulting in a gap between current TTA technology and real-world dubbing demands. Consequently, the mismatch between text descriptions and acoustic scene descriptions poses a significant challenge, primarily due to the technology's limited capability in content planning and detailing for dubbing tasks. Additionally, the absence of a multi-modal dubbing dataset tailored for real-world scenarios is a critical factor hindering technological advancement.

In this paper, we establish a Multi-modal Image and Narrative Text Dubbing Dataset (MINT), aimed at fine-grained foley audio content planning and generation. The dataset encompasses key application scenarios for dubbing tasks, including narrative text descriptions and intricate visual expressions from images, addressing the limitations of existing TTA datasets, which only describe acoustic environments. Based on this dataset, we propose a Foley Audio Content Planning, Generation, and Alignment (CPGA) framework. To address the limitations of existing TTA technology in understanding and planning complex prompts, we have developed a Foley Audio Content Planning module, leveraging large language models (LLMs) to enhance comprehension of complex texts and cross-modal prompts. Furthermore, the training process is optimized by employing Proximal Policy Optimization (PPO) based reinforcement learning. Experimental results demonstrate that the proposed MINT dataset significantly enhances multi-modal dubbing tasks. Compared to models such as GPT-4, our framework excels in supporting general foley audio generation tasks, markedly improving both the alignment and the auditory realism of the generated audio. The main contributions of this article include the following three parts:
\begin{itemize}[leftmargin=20pt]

\item \textbf{MINT dataset:}The MINT dataset was constructed, representing, to our best knowledge, the first dataset for multi-modal Foley audio dubbing involving long narrative texts and images. Long narrative texts present challenges due to information redundancy, conflicts, and interference, making direct application in Foley audio description difficult. Similarly, image content descriptions often deviate from the corresponding Foley audio descriptions across modalities. The establishment of this dataset significantly advances content planning tasks based on complex multi-modal promotes. Experiments demonstrate that the proposed MINT dataset effectively enhances the performance of dubbing tasks in real-world scenarios.

\item \textbf{CPGA framework:}The foley audio Content Planning, Generation, and Alignment (CPGA) framework is proposed, capable of generating foley audio that matches complex multi-modal promotes. The content planning module leverages the large model's deep understanding of multi-modal information, fuses and analyzes these promotes to produce accurate textual descriptions, addressing the time-consuming manual design and adjustment issues inherent in existing technologies. Experiments confirm the proposed method's enhancements in the generalization and robustness of Foley audio generation.

\item \textbf{Acoustic guidance PPO:} Compared with the traditional method, which only optimizes foley audio content planning module and generation module respectively, the proposed direct end-to-end acoustic guidance Proximal Policy Optimization avoids cumulative error loss. Experiments show that the proposed method effectively improves the generation accuracy of Foley audio.
\end{itemize}

\section{MINT: Multi-modal Image and Narrative Text Dubbing Dataset} \label{data}

\subsection{Motivation}
Our motivation for constructing this dataset is based on the following three points.

\begin{itemize}[itemsep=-0.5ex,topsep=-1ex, partopsep=-1ex] 
\item Expanding Applications and Technological Versatility: Text-to-Audio (TTA) techniques primarily focus on the precise acoustic description of Foley audio. The annotation space formed by these prompts is relatively small compared to open scenario description spaces. Although this makes the conversion from prompt to audio relatively straightforward, it also limits the current application scenarios. Despite Foley sound's critical role in post-production, there is no comprehensive dataset for Foley sound generation. To enhance TTA models' ability to generate high-quality Foley sounds, a multimodal input dataset is urgently needed.
\item Enhancing Technological Robustness: Mainstream TTA models are based on diffusion model architecture, where the quality of generation is highly correlated with the input prompt. This necessitates designers to extract precise prompts from multimodal ambiguous prompts in film and television dubbing. To address this, our dataset includes complex multimodal descriptions as inputs and concise, accurate prompts as targets, thereby reducing the burden of prompt engineering for TTA models.
\item Promoting the Development of Audio Content planning and Creation: Current audio generation research focuses on presenting given content under clear constraints, without the need to plan the details involved in the audio content. Traditional audio generation tasks primarily emphasize modality-aligned conversion. The establishment of this dataset can facilitate research on cross-modal conversion in non-aligned scenarios, emphasizing understanding and creation.
\end{itemize}
\subsection{Guidelines}

In this section, we introduce the principles and methods for constructing the dataset.

\begin{itemize}[itemsep=-0.5ex,topsep=-1ex, partopsep=-1ex] 
    \item Simulating Realistic and Robust Scenarios: To better simulate real-world dubbing and enhance the robustness of TTA models, we constructed a multimodal dubbing dataset using filtered AudioCaps\citep{kim2019audiocaps} data. AudioCaps, derived from AudioSet\citep{gemmeke2017audio}, broadly covers natural audio. To align with real-world dubbing demands, we augmented AudioCaps with images and extended textual descriptions using advanced language models. Each dataset entry now includes an image, extended text, original audio caption, and audio, forming quadruplets. These components can be used in various combinations to facilitate modality alignment and simulate scenarios with missing modalities. This ensures the dataset better reflects real-world tasks, improving model performance in practical applications.
    \item Diverse Generation: To ensure the generalization ability of the constructed dataset, we employed various generation and screening methods, leveraging human expertise, advanced language model capabilities, and extended narrative text evaluation techniques. This comprehensive approach guarantees the diversity and generalizability of the dataset.

\end{itemize}

\subsection{Dataset construction process}

\textbf{Image modal part of MINT dataset.} We randomly selected a series of frames from YouTube videos corresponding to AudioCaps. Subsequently, we employed image clustering techniques to identify the single frame that best represents the video's content. Finally We use manual evaluation to filter out images that are blurry or unclear in meaning.

\begin{figure*}[t]
    \centering
    \includegraphics[width=\textwidth]{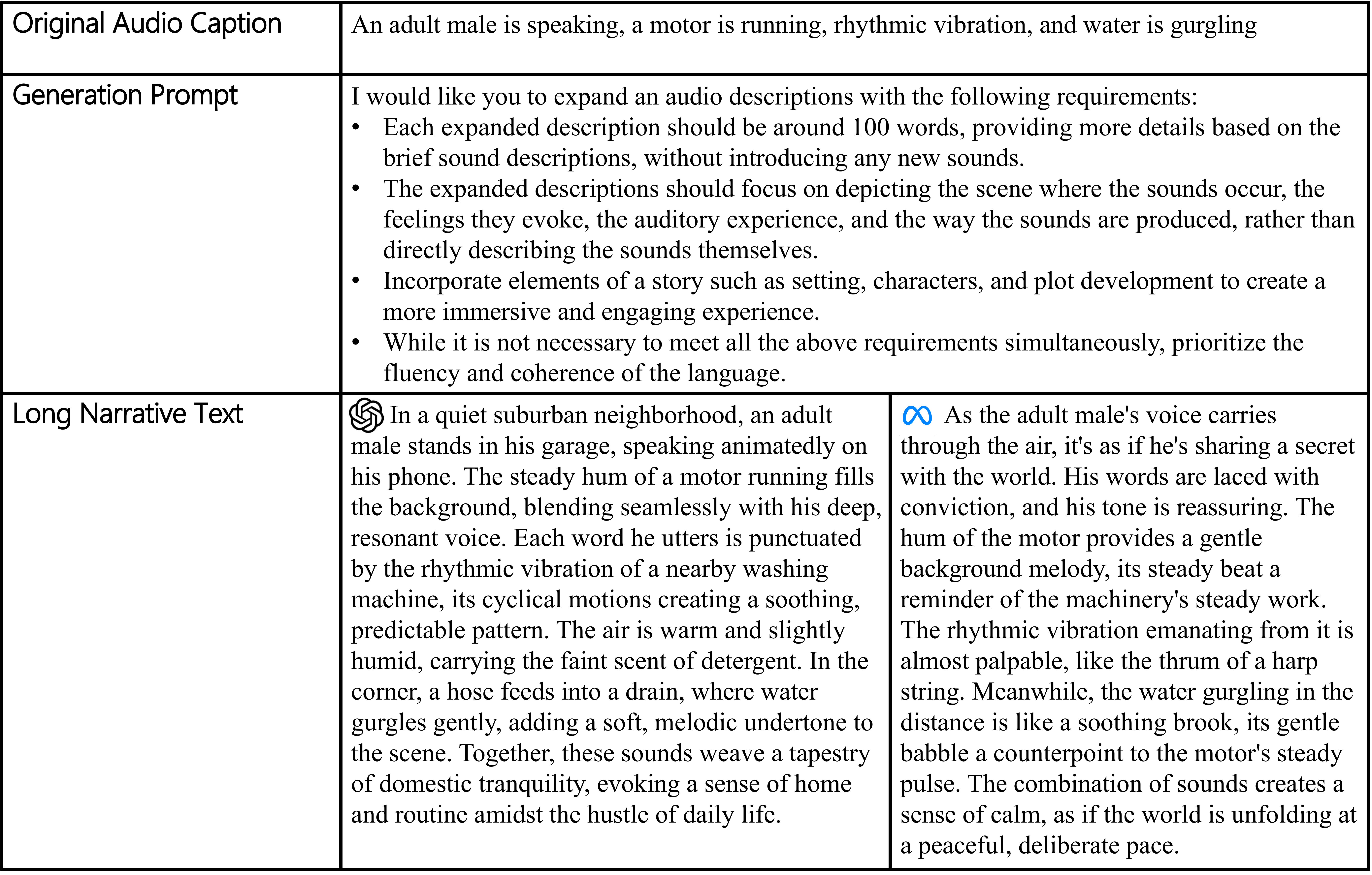}
    \caption{LLM generation example.}
    \label{fig:exp}
\end{figure*}

\textbf{Text modal part of MINT dataset.} Based on the original AudioCaps audio descriptions, we utilize Large Language Models (LMMs) to expand them into long narrtive texts. Our prompts explicitly require the models to follow specific rules for expansion:  \textbf{ (1) No addition of new acoustic elements. (2) Setting the narrative text's plot and characters. (3) Expanding on the subjective auditory experiences. } More details are shown in Figure \ref{fig:exp}.

To ensure the diversity of narrative texts reflective of the real world, we employ multiple mainstream LMMs, including GPT-4 \citep{openai2024gpt4}, LLAMA3-8B, and LLAMA3-70B\citep{llama3modelcard}, to generate different narrative texts. This approach simulates the broad quality range of real-world open-domain narrative text. All generated texts are then scored through human evaluation, and low-scoring texts are filtered out.

\subsection{Overall Statistics}
The expanded dataset comprises 35,363 training samples, 2,532 validation samples, and 2,121 test samples. To verify the comprehensiveness of the dubbing scenarios, we categorized common dubbing scenes into four major categories and eleven subcategories:
\begin{itemize}[itemsep=-0.5ex,topsep=-1ex, partopsep=-1ex] 
\item Natural Environmental Sounds (Weather sounds, Water sounds, Animal sounds)
\item Urban Environmental Sounds (Traffic sounds, Crowd sounds, Construction sounds)
\item Indoor Environmental Sounds (Household appliance sounds, Office environment sounds, Daily household sounds)
\item Industrial Environmental Sounds (Factory machine sounds, Tool usage sounds)
\end{itemize}
We counted the occurrences of all scene sounds within the dataset. Given that the audio in AudioCaps often includes multiple mixed sounds, a single data sample may contain multiple categories of scene sounds. The statistical results indicate that our dataset extensively covers real-world dubbing scenarios. The results are shown in Table \ref{tab:count}. 

\begin{table}[]
\captionsetup{skip=10pt}
\resizebox{\textwidth}{!}{%
\begin{tabular}{|c|ccc|ccl|}
\hline
\textbf{Environmental} & \multicolumn{3}{c|}{\textbf{Natural}}                                                                                           & \multicolumn{3}{c|}{\textbf{Urban}}                                                                         \\ \hline
\textbf{Subset Sounds} & \multicolumn{1}{c|}{\textbf{Weather}}             & \multicolumn{1}{c|}{\textbf{Water}}              & \textbf{Animal}          & \multicolumn{1}{c|}{\textbf{Traffic}}         & \multicolumn{1}{c|}{\textbf{Crowd}} & \textbf{Consturction} \\ \hline
\textbf{Count}         & \multicolumn{1}{c|}{9439}                         & \multicolumn{1}{c|}{9438}                        & 9316                     & \multicolumn{1}{c|}{9379}                     & \multicolumn{1}{c|}{8222}           & \multicolumn{1}{c|}{3111}                  \\ \hline
\textbf{Environmental} & \multicolumn{3}{c|}{\textbf{Indoor}}                                                                                            & \multicolumn{3}{c|}{\textbf{Industrial}}                                                                    \\ \hline
\textbf{Subset Sounds} & \multicolumn{1}{c|}{\textbf{Household appliance}} & \multicolumn{1}{c|}{\textbf{Office environment}} & \textbf{Daily household} & \multicolumn{1}{c|}{\textbf{Factory machine}} & \multicolumn{2}{c|}{\textbf{Factory machine}}               \\ \hline
\textbf{Count}         & \multicolumn{1}{c|}{9514}                         & \multicolumn{1}{c|}{6669}                        & 9763                     & \multicolumn{1}{c|}{8526}                     & \multicolumn{2}{c|}{4352}                                   \\ \hline
\end{tabular}%
}
\caption{Count of Scene Category in the Dataset}
\label{tab:count}
\end{table}

\section{CPGA:Foley audio Content Planning-Generation-Alignment framework}

\begin{figure*}[t]
    \centering
    \includegraphics[width=1.0\textwidth]{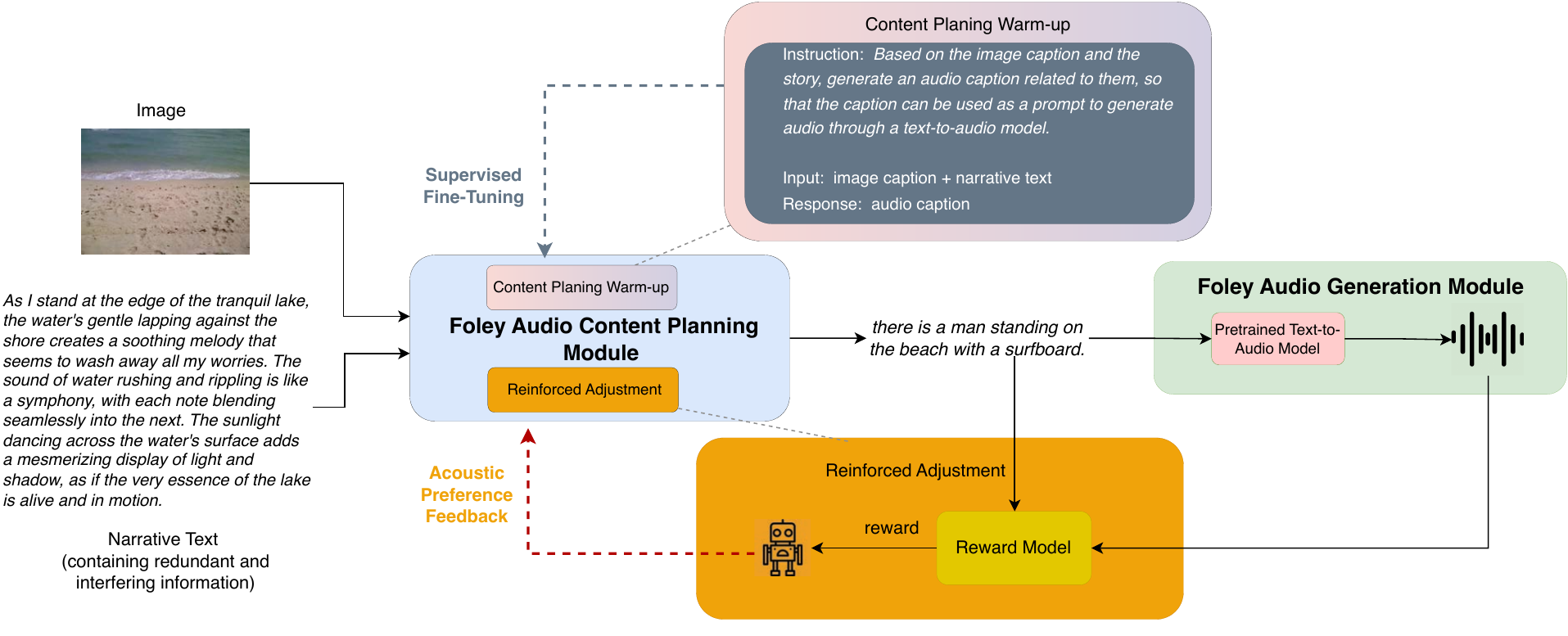}
    \caption{The overall framework of our proposed method.}
    \label{fig:model_arc}
\end{figure*}

The real-world dubbing scenario encompasses a rich tapestry of multimodal inputs, including both visual elements (images or video) and narrative text. Real-world dubbing scenarios necessitate a multimodal approach to audio generation.  This involves not just simple audio generation but a nuanced understanding and integration of multimodal cues—something that conventional TTA technologies are ill-equipped to handle. To bridge the gap between the requirements of real-world dubbing scenarios and the capabilities of TTA technologies, we propose the incorporation of a Foley audio Content Planning, Generation, and Alignment framework (CPGA) into the foley audio generation process. The CPGA  is designed to comprehend and analyze the complexities of multimodal inputs, addressing issues such as redundancies, noise, weak conflicts, and omissions that are prevalent in real-world scenarios. By leveraging the deep understanding capabilities of LMMs, the Foley audio Content Planning module formulates a detailed content plan that guides the foley audio generation process, ensuring that the generated audio is in harmony with both the visual and narrative aspects of the content. Upon establishing a comprehensive Foley audio content plan, the framework proceeds with audio generation using a modified TTA model that is informed by the insights from the Foley audio Content Planning module. This ensures that the generated audio is not only contextually accurate but also enriched with the nuances captured in the content plan. To ensure optimal audio output quality, the framework incorporates an alignment module that employs reinforcement learning mechanisms. This module refines the audio generation process by establishing a reward function based on the effectiveness of the generated audio.

\subsection{Foley Audio Content Planning}
The Foley Audio Content Planning module is a cornerstone in bridging the existing gap between the capabilities of large generative models in audio generation and the nuanced requirements of real-world dubbing applications. This module capitalizes on the recent successes of large language models in various Natural Language Processing  tasks. With their profound contextual understanding, honed by extensive training data, large language models have excelled at tasks such as Text Summarization, Text Simplification, and Paraphrase Generation. Inspired by these capabilities, we harness large language models to explicitly fuse the conditional information required for multimodal dubbing tasks.

The initial step in our Foley audio Content Planning module involves the use of pre-trained image caption models to extract visual information from images, which could be static pictures or keyframes from videos. These models translate the visual data into textual descriptions, effectively converting visual cues into a modality that can be processed alongside textual data. Once the visual information is converted into text, it is concatenated with the corresponding narrative text to form an extended text sequence. The narrative text typically consists of story descriptions that provide context and direction for the dubbing process. However, these descriptions may contain noise and superfluous details that pose challenges for the TTA systems. To refine the content planning process, we employ large language models to perform information fusion on the combined text sequence. Through the use of specific instructions, the language model is directed to sift through the extended text, identifying and summarizing information that positively contributes to the generation of sound effects. The Foley audio Content Planning module constructs input prompts that are structured to facilitate the language model's task. The prompts are composed as follows:
\textbf{<Instruction><Image \hspace{3pt}Caption><Narrative \hspace{3pt}Text>}. By following this structured input, the language model is able to focus on the relevant aspects of the text sequence, filtering out irrelevant details and emphasizing information that aids in the creation of Foley audio content.

\subsection{Foley Audio Generation}

Following the successful integration of visual and narrative information into a concise text sequence via the proposed information fusion approach in the first step, we advance to the second step: audio generation using a TTA model. Among various TTA models, those based on latent diffusion models have shown remarkable capabilities in generating high-quality audio that aligns closely with the text inputs. A standout example in this domain is Tango, which utilizes the Flan-T5, a large language model, as its text encoder. Flan-T5 has been pre-trained on a vast array of chain-of-thought and instruction-based datasets, endowing it with exceptional generalization performance across different NLP tasks. This makes it an ideal choice for encoding the relatively accurate audio caption derived from step one for Foley audio generation.

Although the integration of the Foley audio Content Planning and the use of Tango for Foley audio generation represent significant advancements, relying solely on the Foley Audio Content Planning could lead to cumulative errors throughout the pipeline. This scenario essentially represents an open-loop control system, where a gap remains between the generated text prompts and the final audio output, potentially affecting the controllability of the generation process.

\subsection{End to end alignment based on acoustic guidance Proximal Policy Optimization}
To guarantee optimal outcomes in the final generated audio, our framework incorporates an alignment module that utilizes a reinforcement learning mechanism for fine-tuning the language model. This step is crucial for enhancing the precision of the prompts generated under the instruction of the language model, which, in turn, ensures that the audio produced is of the highest quality.
The reinforcement learning mechanism leverages real-world audio samples to imbue the large language model with a richer understanding of acoustic information. This is achieved by setting up a reward function that evaluates the acoustic similarity between the generated audio and authentic audio samples. By optimizing towards this reward function, the large language model becomes adept at aligning textual prompts with the corresponding acoustic properties required for the audio output. This alignment is essential for maintaining congruence between the generated audio and the multimodal inputs, which includes visual cues and narrative text.

Proximal Policy Optimization (PPO) serves as the method of choice within this reinforcement learning framework to accomplish end-to-end alignment. PPO directly tunes the model's parameters by maximizing the reward function that reflects the quality of the audio output. This optimization process focuses on the end result, streamlining the alignment of the generated audio with the expected acoustic profile. As the model iterates through this process, it learns to produce prompts that lead to the most immersive and accurate Foley audio.
Through the application of this reinforcement learning strategy, our framework achieves a robust and efficient pipeline capable of generating Foley audio that is not just contextually and thematically consistent with the visual and narrative elements but also acoustically rich and authentic. This end-to-end optimization ensures that the auditory content not only meets but exceeds the standards required for real-world applications, thereby providing an Enhanced immersive experience.

\section{Experiments and Analysis}

\subsection{Dataset}
In this work, we use the proposed MINT dataset to construct the training and testing sets used in the experiments. The training set contains a total of 35,363 entries, while the testing set contains 2,532 entries. In the multimodal dubbing generation method proposed in this paper, the information carried by an image is expressed by converting it into an image caption. Therefore, it is necessary to preprocess the images in the dataset to improve training speed. To achieve this, we use the BLIP \citep{blip}, pretrained on the COCO \citep{coco} dataset, to extract image captions. During the preprocessing stage, for each image, we use BLIP to generate 5 image captions. Then, we use the CLIP \citep{clip} to calculate the similarity between the image and each of the 5 image captions, selecting the caption with the highest CLIP score as the final choice.

\subsection{Training Setting}
\label{resources}
We use GPT-2 (GPT2-124M) \citep{radford2019gpt-2} as the model for training. The Stage \uppercase\expandafter{\romannumeral1} Content Planing Warm-up is to train the language model in content planning, enabling it to correctly extract valid information from long texts that contain complex and redundant information. The Stage \uppercase\expandafter{\romannumeral2} Reinforced Adjustment involves using the reinforcement learning algorithm PPO to feed acoustic information from the audio back into the language model, allowing the language model to generate better prompt texts for the TTA model.

\textbf{Stage \uppercase\expandafter{\romannumeral1}: Content Planing Warm-up}
Training the language model to learn content planning requires a high-quality dataset. We consider the textual descriptions in the AudioCaps dataset to be high-quality audio captions, as they are manually annotated and have undergone strict screening. We construct a high-quality dataset to train GPT-2, consisting of <query, response> pairs. The query is a combination of Instruction, image caption, and narrative text, while the response is the corresponding textual description from AudioCaps. We train for 50 epochs on a single 4090 GPU with a batch size of 16, and the training duration is 4 hours.

\textbf{Stage \uppercase\expandafter{\romannumeral2}: Reinforced Adjustment}
In the reinforcement learning adjustment phase, we randomly select a data entry from the training set and concatenate the image caption, narrative text, and Instruction from it, then input this concatenated text into the GPT-2 model fine-tuned in Step 1. After content planning by GPT-2, it generates a more precise textual description that matches the input distribution of the TTA model. Subsequently, we use CLAP \citep{wu2023clap} to extract features from both the textual description and the audio from the selected data and calculate their similarity. The calculated similarity serves as the reward for the PPO algorithm, which is then used to fine-tune the GPT-2 model. During this training phase, we use a learning rate of 1e-6, a batch size of 16, a mini-batch size of 4, and train for 10 epochs on two 4090 GPUs. The training duration is approximately 35 hours.

\subsection{Evaluation Metrics}
\label{Evaluation Metrics}
We perform both objective evaluation and human subjective evaluation on the generated audio. The primary metrics used for objective evaluation include the Frechet Distance (FD) and Kullback-Leibler (KL) divergence. The FD score quantifies the overall similarity between the generated audio samples and the target samples without the need for paired reference audio samples. The KL divergence is calculated using paired samples and measures the divergence between two probability distributions. Both metrics are based on the state-of-the-art audio classifier PANNs \citep{kong2020panns}. For subjective evaluation, following previous methods in the TTA field \citep{kreuk2022audiogen, liu2023audioldm}, we ask five human evaluators to assess two aspects of the generated audio: overall audio quality (OVL) and relevance to the text caption (REL). We randomly select 20 audio samples generated by each method and ask participants to rate them on a scale from 1 to 100.

\begin{table*}[ht!]
\centering
\caption{The subjective evaluation results of different TTA models and vision-language models (VLMs). In this comparison of different VLMs , Tango is used to generate audio from the audio captions.}
\resizebox{1.0\textwidth}{!}{
\begin{tabular}{cc|cc|cc|cc}
\hline
\multirow{2}{*}{\textbf{Model}} & \multirow{2}{*}{\textbf{Input}} & \multicolumn{2}{c|}{\textbf{Subjective Metrics}} & \multirow{2}{*}{\textbf{Model}} & \multirow{2}{*}{\textbf{Input}}            & \multicolumn{2}{c}{\textbf{Subjective Metrics}}                   \\
                                &                                 & OVL~$\uparrow$                     & REL~$\uparrow$                    &                                 &                                            & OVL~$\uparrow$                             & REL~$\uparrow$                             \\ \hline
GroundTruth                              & -                               & 87.56                   & 85.72                  & GroundTruth                              & -                                          & 87.56                           & 85.72                           \\ \hline
AudioGen \citep{kreuk2022audiogen}                        & narrative text                  & 37.05                   & 29.66                  & \multirow{2}{*}{Moondream2}     & \multirow{2}{*}{image \&\& narrative text} & \multirow{2}{*}{50.56}          & \multirow{2}{*}{47.52}          \\
AudioLDM \citep{liu2023audioldm}                       & narrative text                  & 35.11                   & 30.22                  &                                 &                                            &                                 &                                 \\
Tango \citep{majumder2024tango}                           & narrative text                  & 41.36                   & 32.15                  & \multirow{2}{*}{DeepSeek-VL \citep{lu2024deepseekvl}}    & \multirow{2}{*}{image \&\& narrative text}   & \multirow{2}{*}{58.18}          & \multirow{2}{*}{56.67}          \\ \cline{1-4}
AudioGen                        & AudioCaps test text             & 67.52                   & 65.23                  &                                 &                                            &                                 &                                 \\
AudioLDM                        & AudioCaps test text             & 72.35                   & 70.10                  & \multirow{2}{*}{LLaVA \citep{liu2023llava}}          & \multirow{2}{*}{image \&\& narrative text}   & \multirow{2}{*}{\textbf{63.73}} & \multirow{2}{*}{\textbf{60.65}} \\
Tango                           & AudioCaps test text             & \textbf{83.55}          & \textbf{81.29}         &                                 &                                            &                                 &                                 \\ \hline
\end{tabular}
}
\label{tab:compare_tta_vlm}
\end{table*}

\begin{table*}[ht!]
\centering
\caption{The results of our proposed method and other methods in both objective and subjective evaluations. Tango is also used as the audio generation model in this comparison.}
\resizebox{1.0\textwidth}{!}{
\begin{tabular}{cc|cc|cc}
\toprule
\multirow{2}{*}{\textbf{Model}} & \multirow{2}{*}{\textbf{Input Content}} & \multicolumn{2}{c|}{\textbf{Objective Metrics}} & \multicolumn{2}{c}{\textbf{Subjective Metrics}} \\ 
 & & FD~$\downarrow$ & KL~$\downarrow$  & OVL~$\uparrow$ & REL~$\uparrow$ \\
\midrule
GroundTruth &$-$  & $-$ & $-$ & $87.56$ & $85.72$ \\
\midrule
GPT-2 &$narrative\hspace{3pt}text$  & $45.09$ & $3.95$ & $45.18$ & $33.67$ \\

GPT-2 &$image\hspace{3pt} caption$  & $48.26$ & $4.00$ & $42.56$ & $31.52$ \\

GPT-2 &$image\hspace{3pt} caption \&\& narrative\hspace{3pt} text$  & $46.40$ & $3.97$ & $48.56$ & $32.17$ \\
\midrule
SFT-GPT-2 &$narrative\hspace{3pt} text$  & $28.87$ & $1.70$ & $65.52$ & $62.23$ \\

SFT-GPT-2 &$image\hspace{3pt} caption$  & $33.78$ & $3.35$ & $62.16$ & $59.31$ \\

SFT-GPT-2 &$image\hspace{3pt} caption \&\& narrative\hspace{3pt} text$  & $27.09$ & $\mathbf{1.59}$ & $73.55$ & $71.29$ \\
\midrule
LLaVA &$image\&\& narrative\hspace{3pt} text$  & $30.28$ & $1.82$ & $63.73$ & $60.65$ \\
\midrule
PPO-SFT-GPT-2 & $image\hspace{3pt} caption \&\& narrative\hspace{3pt} text$  & $\mathbf{26.99}$ & $1.66$ & $\mathbf{78.35}$ & $\mathbf{75.96}$ \\
\bottomrule
\end{tabular}
}
\label{tab:ResultsWithTango}
\end{table*}

\subsection{MINT dataset Tests on mainstream TTA models}

In this section, we directly tested the test set of the MINT dataset on the current mainstream TTA models. Observations from the left side of the Table \ref{tab:compare_tta_vlm} indicate that when only the narrative text input from the MINT dataset is considered, the existing mainstream TTA models (including AudioGen, AudioLDM, Tango) fail to effectively process the narrative text. When compared to the annotated audio text descriptions in AudioCaps, there is a noticeable drop in performance, with even the top-performing TANGO model suffering a performance loss of approximately 40\% on both OVL and REL metrics. The primary reason is that the current TTA models have a limited capacity to comprehend the existing text descriptions, and their functionality is confined to reconstructing precise acoustic descriptions. This experiment further confirms the significant modeling challenges in simulating the real-world application scenarios of the MINT dataset that we have constructed.
\subsection{MINT dataset Tests on mainstream Multi-modal Large Models}

In this section, we employ the test set from the constructed MINT dataset to evaluate several leading multimodal large models. To ensure optimal performance of the MLMs, we meticulously selected the most effective test results as the evaluation set for each sample. The experimental results on the right side of Table \ref{tab:compare_tta_vlm} reveal that, for multi-modal prompt inputs, LLaVA outperforms Moondream2\footnote{\url{https://huggingface.co/vikhyatk/moondream2}} and DeepSeek-VL in multi-modal foley dubbing. Integrating image modal guidance, LLaVA also performs better than the traditional TTA model, yet it still significantly trails the GroundTruth. Our analysis indicates that while multimodal large models possess certain capabilities in processing multimodal information and comprehension, they lack specialized training for the dubbing task, resulting in notable shortcomings in this area of expertise.

\subsection{Evaluation on proposed CGPA framework}

In this section, due to the limited computing resources, we selected gpt-2 as the basic model of content planning module to evaluate the CPGA framework we proposed, and evaluated it with FD, KL, OVL and rel. It can be seen from Table \ref{tab:ResultsWithTango} that the performance of Foley audio dubbing has been significantly improved after SFT operation compared with that without SFT operation due to full learning of the mode of dubbing task. At the same time, it can be observed that the overall performance is better when multimodal prompts are used as input. We speculate that the model learning can effectively promote the model's in-depth understanding of multimodal dubbing tasks. In addition, it can also be observed that the subjective and objective indicators of PPO-SFT-GPT-2 proposed by us are better than those of LLaVA, and the end to end alignment based on acoustic guidance PPO method also further improves the performance, reaching an overall better performance for the current multimodal dubbing task.

\section{Conclusions and Future Work}

In this paper, we introduce the MINT dataset, a multi-modal dataset designed for foley audio dubbing tasks involving narrative texts and images. 
We also propose the CPGA framework, which integrates content planning, generation, and alignment for foley audio. This framework leverages the deep understanding capabilities of large models to process and analyze multi-modal prompts, significantly reducing the reliance on manual design and adjustment. The CPGA framework enhances the generalization and robustness of foley audio dubbing, ensuring that the generated audio aligns well with the provided multi-modal prompts.
Furthermore, we employ a reinforcement learning approach based on PPO with acoustic guidance. This end-to-end optimization method avoids the cumulative loss of errors that can occur when optimizing individual modules separately. 
Experiments have demonstrated that the proposed CPGA framework significantly enhances dubbing tasks. The content planning module effectively addresses the current challenges of multimodal prompt fusion and long narrative text comprehension. Even when utilizing the relatively lightweight GPT-2 model, our framework outperforms open-source multimodal large models such as LLaVA, DeepSeek-VL, and Moondream2. Future work focuses on further improve the content key element planning and research on dubbing in the case of more modals.

\begin{ack}
This work is supported by the National Natural Sciencel Foundation of China (NSFC) (No.62101553, No.62306316, No.U21B20210, No.62201571).
\end{ack}

\bibliographystyle{plainnat}

\newpage

\section{Appendix}
\definecolor{mycolor}{RGB}{255,165,0}

\subsection{Data Format}
Our proposed dataset can be accessed from the following link.\url{https://github.com/borisfrb/MINT}

Our dataset includes images, narrative text, audio captions, and audio. We organize the data using JSON files, where each line represents a data sample. Table \ref{example} is an example of a data sample.
You can use the yt-dlp\footnote{\url{https://github.com/yt-dlp/yt-dlp}} to retrieve audio files using the provided youtube\_id and audio\_start\_time. Note that when using the yt-dlp tool to fetch audio files, ensure that the duration for fetching audio is set to 10 seconds, as specified by AudioSet and AudioCaps. For images, we provide their indices in the JSON file. The actual images can be downloaded from Zenodo, and the download links can be obtained from the provided link.
\begin{table*}[ht!]
\centering
\caption{Example of a data sample}
\resizebox{\textwidth}{!}{
\begin{tabular}{|l|}
\hline
\{"audiocaps\_id": "97151", \\"youtube\_id": "vfY\_TJq7n\_U", \\"audio\_start\_time": "130", \\"audio\_caption": "Rustling occurs, ducks quack and water splashes, followed by an adult female and adult male\\ speaking and duck calls being blown", \\"image": "97151.png", \\"narrative\_text": "As I make my way along the winding path, I come across a loving couple, their gentle conversation\\ a warm and intimate accompaniment to the natural soundscape. The adult female's voice is soft and melodious, while \\the adult male's is deep and soothing. Their words are lost in the distance, but the love and contentment in their tone is \\palpable. Suddenly, a duck call pierces the air, followed by a chorus of quacks and honks from the ducks in the water. \\The sounds blend together in perfect harmony, a beautiful tapestry of sound that envelops me in its serenity."\} \\ \hline
\end{tabular}
}
\label{example}
\end{table*}

\subsection{Details of Subjective Evaluation}
\subsubsection{Evaluation metrics}
Overall audio quality (OVL) and relevance to the text caption (REL) are subjective measures used to evaluate the effectiveness of audio generated by text-to-audio models. Table \ref{ovl_detail} and Table \ref{rel_detail} show the scoring criteria we designed.

\begin{table*}[ht!]
\centering
\caption{Overall audio quality rating level description}
\resizebox{\textwidth}{!}{
\begin{tabular}{|c|l|}
\hline
OVL      & \multicolumn{1}{c|}{Level Description}                                                                                                                                                                                       \\ \hline
80-100   & \begin{tabular}[c]{@{}l@{}}The audio quality is extremely high, the sound is clear and natural, with almost no noise or distortion, \\ and the overall performance is very close to real environment recording.\end{tabular} \\ \hline
70-89    & \begin{tabular}[c]{@{}l@{}}The audio quality is high, the sound is clear, and there may be slight noise or distortion occasionally, \\ but it does not affect the overall listening experience.\end{tabular}                 \\ \hline
50-69    & \begin{tabular}[c]{@{}l@{}}The audio quality is moderate and the sound is basically clear, but there is obvious noise or distortion \\ that may affect the listening experience.\end{tabular}                                \\ \hline
30-49    & \begin{tabular}[c]{@{}l@{}}The audio quality is low, and there is significant noise or distortion in the sound, which seriously affects \\ the overall listening experience.\end{tabular}                                    \\ \hline
Below 30 & \begin{tabular}[c]{@{}l@{}}The audio quality is very poor, the sound is blurry, the noise or distortion is very serious, and it is almost \\ impossible to listen normally.\end{tabular}                                     \\ \hline
\end{tabular}
}
\label{ovl_detail}
\end{table*}

\begin{table*}[ht!]
\centering
\caption{Relevance to the text caption rating level description}
\resizebox{\textwidth}{!}{
\begin{tabular}{|c|l|}
\hline
REL      & \multicolumn{1}{c|}{Level Description}                                                                                                                                                                                        \\ \hline
80-100   & \begin{tabular}[c]{@{}l@{}}The audio and text descriptions are highly consistent, with every detail perfectly matching the text, \\ and the auditory content almost perfectly matches the text content.\end{tabular}          \\ \hline
70-89    & \begin{tabular}[c]{@{}l@{}}The audio and text descriptions are relatively consistent, and the main content and details are basically \\ consistent with the text. Only a few details may have slight deviations.\end{tabular} \\ \hline
50-69    & \begin{tabular}[c]{@{}l@{}}The audio and text descriptions are consistent and can reflect the main content of the text, but there are \\ obvious deviations or missing details.\end{tabular}                                  \\ \hline
30-49    & \begin{tabular}[c]{@{}l@{}}The similarity between audio and text descriptions is low, with only some content matching the text desc-\\ ription and most details being inaccurate or incorrect.\end{tabular}                   \\ \hline
Below 30 & \begin{tabular}[c]{@{}l@{}}The audio and text descriptions are almost inconsistent, with significant differences in content that cannot \\ reflect the main information of the text.\end{tabular}                             \\ \hline
\end{tabular}
}
\label{rel_detail}
\end{table*}

\subsubsection{Evaluators Selection}
Participants for the OVL and REL listening tests should be selected based on specific criteria to ensure they represent the target audience of the generated audio. Typically, participants should have normal hearing capabilities and be unfamiliar with the specific work of this study to avoid bias. Ensuring diversity among paid participants helps achieve a more generalizable assessment of audio quality.

\textbf{Pre-test Training.} To ensure scoring consistency, evaluators undergo comprehensive training before the actual test. This training includes:

\begin{enumerate}
    \item \textbf{Introduction to the Evaluation Criteria}: Detailed explanation of the OVL and REL scoring criteria, including examples of each rating level.
    \item \textbf{Sample Evaluations}: Evaluators listen to pre-selected audio samples that exemplify different aspects of the OVL and REL scoring criteria. These samples should cover a range of quality levels from poor to excellent.
    \item \textbf{Practice Session}: Evaluators participate in practice sessions where they rate additional samples and receive feedback on their ratings. This step helps calibrate their judgments and align their understanding of the scoring criteria.
\end{enumerate}

\textbf{Consistency Checks.} During the training phase, evaluators' ratings are monitored for consistency. Statistical measures such as inter-rater reliability (e.g., Cronbach’s alpha) can be calculated to ensure agreement among evaluators. Evaluators who exhibit significant deviations from the group consensus may require additional training or be excluded from the final testing.
\subsubsection{Evaluation Scale}

We ask five evaluators to assess the generated audio from two aspects: OVL and REL. We select 20 audio samples generated by each method and ask participants to rate them on a scale from 1 to 100. The scoring table used by each evaluator is shown in Table \ref{score_detail}.

\begin{table*}[ht!]
\centering
\caption{Scoring table for evaluators}
\begin{tabular}{|l|l|l|l|l|}
\hline
    & Audio\_1 & Audio\_2 & ..... & Audio\_20 \\ \hline
OVL &          &          &       &           \\ \hline
REL &          &          &       &           \\ \hline
\end{tabular}
\label{score_detail}
\end{table*}

\subsection{Vision-Language Model Testing Details}
We test several vision-language models using the test set from the proposed MINT dataset. First, we require the vision-language models to generate an audio caption related to the image and narrative text given the image, narrative text, and instruction. This audio caption is then fed into the text-to-audio model to generate audio. Finally, we evaluate the generated audio. The models used for testing and the prompts employed for each different model are shown in Table \ref{vlm_model} and Table \ref{vlm_prompt}, respectively.
\begin{table*}[ht!]
\centering
\caption{The link to the pre-trained vision-language model used.}
\begin{tabular}{|c|c|}
\hline
Model       & Link                                                     \\ \hline
Moondream2  & \url{https://huggingface.co/vikhyatk/moondream2}               \\ \hline
DeepSeek-VL & \url{https://huggingface.co/deepseek-ai/deepseek-vl-7b-chat}   \\ \hline
LLaVA       & \url{https://huggingface.co/llava-hf/llava-v1.6-mistral-7b-hf} \\ \hline
\end{tabular}
\label{vlm_model}
\end{table*}

\begin{table*}[ht!]
\centering
\caption{The prompts used during testing with the pre-trained vision-language model. Orange text indicates escape characters. \textbf{Instruction}: "\textit{Based on the image and the narrative text, generate an audio caption related to them, so that the caption can be used as a prompt to generate audio through a text-to-audio model.}"}
\resizebox{\textwidth}{!}{
\begin{tabular}{|c|c|}
\hline
Model       & Prompt                                                                                                                                                         \\ \hline
Moondream2  & "narrative text:\textcolor{mycolor}{\textbackslash{}n}" + \textbf{narrative text} + "\textcolor{mycolor}{\textbackslash{}n}" + \textbf{Instruction}                                                    \\ \hline
DeepSeek-VL & "<image\_placeholder>\textcolor{mycolor}{\textbackslash{}n}narrative text:\textcolor{mycolor}{\textbackslash{}n}" + \textbf{narrative text} + "\textcolor{mycolor}{\textbackslash{}n}" + \textbf{Instruction}                \\ \hline
LLaVA       & "{[}INST{]} <image>\textcolor{mycolor}{\textbackslash{}n}narrative text:\textcolor{mycolor}{\textbackslash{}n}"+ \textbf{narrative text} + "\textcolor{mycolor}{\textbackslash{}n}" + \textbf{Instruction} + "{[}/INST{]}" \\ \hline
\end{tabular}
}
\label{vlm_prompt}
\end{table*}

\subsection{Dubbing Scene Statistics}
\begin{figure*}[t]
    \centering
    \includegraphics[width=\textwidth]{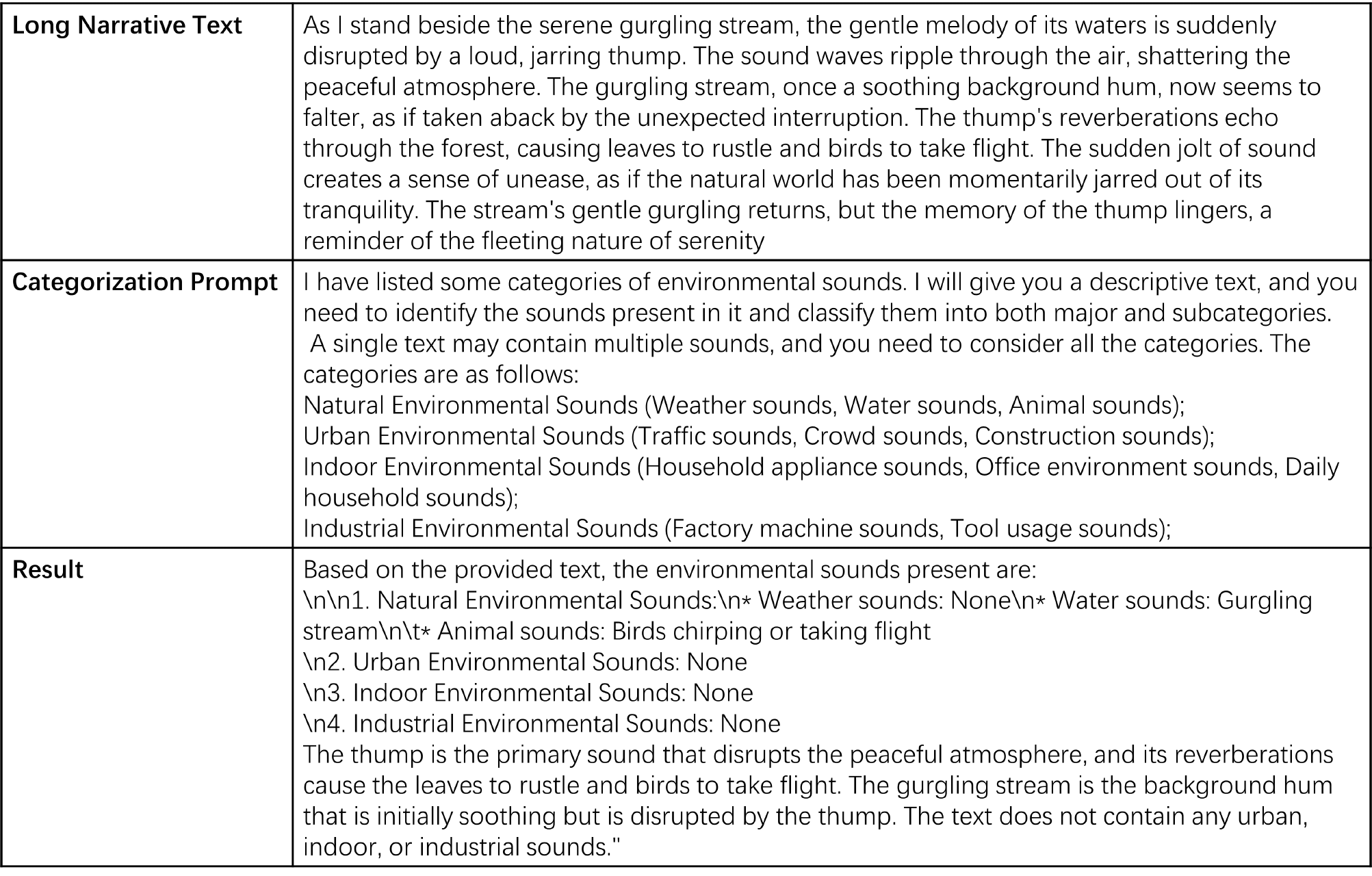}
    \caption{Examples of Scene Classification}
    \label{fig:cls}
\end{figure*}

We employed the LLama-70b model to perform a statistical coverage analysis of dubbing themes within the dataset. The implementation details are presented in the Figure \ref{fig:cls}.

\subsection{Dataset Construction Details}
\begin{figure*}[t]
    \centering
    \includegraphics[width=\textwidth]{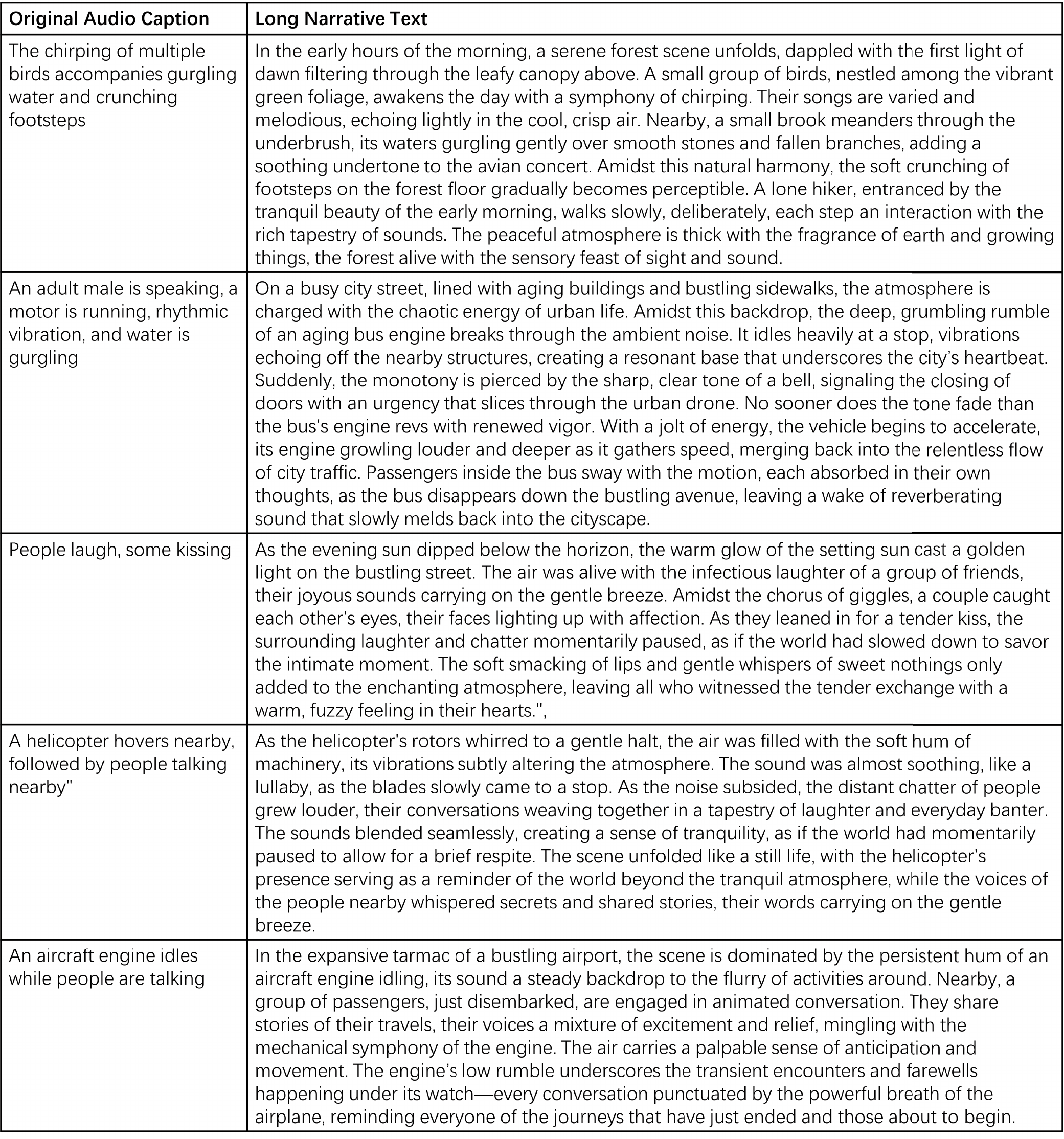}
    \caption{More Examples of Long Narraitive text}
    \label{fig:exp2}
\end{figure*}

To ensure the integrity of the text expansion, we have established several stringent rules. First, the expansion process strictly prohibits the addition of any new auditory elements, ensuring that the narratives are solely based on the original audio content. Second, the models are tasked with constructing complete storylines, setting scenes, and characters, to enhance the narrative's immersion and appeal. Moreover, the expanded content should primarily describe the scenarios depicted by the audio, the emotions evoked, and the manner in which the sounds are produced, rather than directly describing the sounds themselves. Each expanded description, approximately 100 words in length, should elaborate on these elements in greater detail while prioritizing narrative fluency and linguistic coherence. Through this approach, we aim to provide a richer, more storied auditory experience. More results are shown in Figure \ref{fig:exp2}.

\subsection{The Main Claims}

The paper’s contributions and scope are focusing on the creation and characteristics of a new dataset. These sections effectively highlight the dataset’s relevance and potential impact on further research.

\subsection{External Data}

We are utilizing the AudioCaps dataset, which is available under the MIT License. This licensing ensures that we adhere to the required legal standards for use and distribution of the data.

\subsection{Ethics Statement}
The access to this MINT dataset is limited to academic institutions and is for research purposes only. We include frames extracted from each YouTube video as images in the dataset and ensure that these images do not adversely affect the copyright owner's ability to generate revenue from their original content, thereby complying with YouTube's fair use policy. If any copyright owner believes their rights have been infringed, we commit to promptly removing the disputed materials from our dataset.
\end{document}